\documentclass[twocolumn,showpacs,amsmath,amssymb,10pt]{revtex4}
\usepackage{amsfonts}
\usepackage{bm}

\newcommand{\ot}{{\,\otimes\,}}
\newcommand{{\Cd}}{{\mathbb{C}^d}}
\newcommand{\sbsigma}{{\mbox{\scriptsize \boldmath $\sigma$}}}
\newcommand{\sbalpha}{{\mbox{\scriptsize \boldmath $\alpha$}}}
\newcommand{\sbbeta}{{\mbox{\scriptsize \boldmath $\beta$}}}

\newcommand{\ba}{{\mbox{ \boldmath $a$}}}
\newcommand{\bb}{{\mbox{ \boldmath $b$}}}
\newcommand{\sba}{{\mbox{\scriptsize \boldmath $a$}}}

\def\oper{{\mathchoice{\rm 1\mskip-4mu l}{\rm 1\mskip-4mu l}%
{\rm 1\mskip-4.5mu l}{\rm 1\mskip-5mu l}}}
\def\<{\langle}
\def\>{\rangle}

\begin{document}
\title{\textbf{On multipartite invariant states II. \\ Orthogonal symmetry.}} \author{Dariusz
Chru\'sci\'nski and Andrzej Kossakowski\thanks{email:
darch@phys.uni.torun.pl} }
\affiliation{Institute of Physics, Nicolaus Copernicus University,\\
Grudzi\c{a}dzka 5/7, 87--100 Toru\'n, Poland}

\begin{abstract}

We construct a new class of multipartite states possessing
orthogonal symmetry. This new class defines a convex hull of
multipartite states which are invariant under the action of local
unitary operations introduced in our previous paper {\it On
multipartite invariant states I.  Unitary symmetry}. We study
basic properties of multipartite  symmetric states: separability
criteria and multi-PPT conditions.

\end{abstract}

\maketitle
\section{Introduction}

In a recent paper \cite{I} we analyzed  multipartite states
invariant under local unitary operations. For bipartite systems
one has  two classes of unitary invariant  states: Werner states
\cite{Werner1} invariant under
\begin{equation}\label{W}
\rho \ \longrightarrow\  U\ot U \, \rho\, (U \ot U)^\dag\ ,
\end{equation}
for any $U\in U(d)$, where $U(d)$ denotes the group of unitary $d
\times d$ matrices, and isotropic states \cite{Horodecki}  which
are invariant under
\begin{equation}\label{I}
\rho\  \longrightarrow\  U\ot \overline{U} \, \rho \, (U \ot
\overline{U})^\dag\ ,
\end{equation}
where $\overline{U}$ is the complex conjugate of $U$ in some
basis. In \cite{I} we proposed a natural generalization of
bipartite symmetric  states to multipartite systems consisting of
an arbitrary even number of $d$-dimensional subsystems (qudits).

In the present paper we  introduce a new class of states  which
 combines above symmetries (\ref{W}) and (\ref{I}), i.e. it
 contains states which are both $U\ot U$ and $U \ot \overline{U}$--invariant,
 that is, states invariant under all unitary operations $U$ such that $U =
\overline{U}$:
\begin{equation}\label{O}
\rho \ \longrightarrow\  O\ot O \, \rho\, (O \ot O)^T\ ,
\end{equation}
with $O\in O(d) \subset U(d)$, where $O(d)$ denotes the set of $d
\times d$ orthogonal matrices. Such states were first considered
in \cite{Werner2} (see also \cite{Werner3}). In a slightly
different context symmetric states were studied also in
\cite{Plenio}. Recently \cite{Hall} unitary invariant 3-partite
states were used to test multipartite separability criteria.

 Here we present a general construction of
$O \ot O$--invariant states for multipartite systems consisting of
an arbitrary even number of $d$-dimensional subsystems. It turns
out that orthogonally invariant states of $2K$--partite system
(with $K$ being a positive integer) define $(3^{K}-1)$--invariant
simplex. We analyze (multi)separability criteria and the hierarchy
of multi-PPT conditions \cite{I,Peres,PPT}. It is hoped that these
new family would serve as a useful laboratory to study
multipartite entanglement
\cite{MULTI1,MULTI2,MULTI3,MULTI4,MULTI5,MULTI6}.

\section{Bipartite states}
\label{2-PARTIES}

\subsection{Simplex of orthogonally invariant states}

 Let us consider a bipartite Alice--Bob system living in
$\mathcal{H}_{AB} = \mathcal{H}_A \ot \mathcal{H}_B =
(\mathbb{C}^d)^{\ot 2}$. Recall that the space of $U\ot
U$--invariant hermitian operators in $\mathcal{H}_{AB}$ is spanned
by two orthogonal projectors
\begin{equation}\label{Q}
    Q^0 = \frac 12 ( I^{\ot 2}  + \mathbf{F}) \ , \ \ \
    Q^1 = \frac 12 ( I^{\ot 2}  - \mathbf{F})\ ,
\end{equation}
where ${\bf F}$ is a flip operator, i.e.  $\mathbf{F}(\psi\ot
\varphi) = \varphi \ot \psi$, defined by
\begin{equation}\label{}
    \mathbf{F} = \sum_{i,j=1}^d\, |ij\>\<ji| \ .
\end{equation}
In particular this 2-dimensional space contains a line of
normalized (i.e. with unit trace) operators:
\begin{equation}\label{}
    %\mathcal{W}_{\bf q} =
L: \ \ \ (1-q)\,
    \widetilde{Q}^0 + q\,\widetilde{Q}^1 \ ,
\end{equation}
with  $q\in \mathbb{R}$,  and throughout the paper $\widetilde{A}$
stands for $A/\mbox{Tr}A$.  A segment of $L$ with vertices
$\widetilde{Q}^0$ and $\widetilde{Q}^1$ defines a family of
bipartite Werner states:
\begin{equation}\label{}
    \mathcal{W}_{\bf q} =
q_0\, \widetilde{Q}^0 + q_1\,\widetilde{Q}^1 \ ,
\end{equation}
with $q_\alpha\geq 0$, and $q_0 + q_1=1$.

Now, a partial transposition $\oper \ot \tau$ sends points of $L$
into another line $L_\tau=(\oper \ot \tau)L$:
\begin{equation}\label{}
    %\mathcal{W}_{\bf q} =
L_\tau: \ \ \ (1-p)\,
    \widetilde{P}^0 + p\,\widetilde{P}^1 \ ,
\end{equation}
with  $p\in \mathbb{R}$, and $P^\alpha$ denote the following
orthogonal projectors:
\begin{equation}\label{P-alpha}
    P^1 = P^+_d\ , \ \ \ P^0 = I^{\ot 2} - P^1\ ,
\end{equation}
with $P^+_d$ being a 1-dimensional projector corresponding to a
canonical maximally entangled state in $\Cd\ot \Cd$:
\begin{equation}\label{}
    P^+_d = \frac 1d\, (\oper \ot \tau){\bf F} = \frac 1d\, \sum_{i,j=1}^d\, |ii\>\<jj|\ .
\end{equation}
A segment of $L_\tau$ with vertices $\widetilde{P}^0$ and
$\widetilde{P}^1$ defines a family of bipartite isotropic states:
\begin{equation}\label{}
    \mathcal{I}_{\bf p} =
p_0\, \widetilde{P}^0 + p_1\,\widetilde{P}^1 \ ,
\end{equation}
with $p_\alpha\geq 0$, and $p_0 + p_1=1$.

Let us introduce a new class $\Sigma_1$ of bipartite states which
are both $U\ot U$ and $U \ot \overline{U}$--invariant for all $U
\in U(d)$ such that $U = \overline{U}$. Such $U$'s represent real
orthogonal matrices in $O(d)$. Hence, $\Sigma_1$ defines a new
family of symmetric $O \ot O$--invariant states:
\begin{equation}\label{OO}
\rho \ \longrightarrow\  O\ot O \, \rho\, (O \ot O)^T\ ,
\end{equation}
with $O\in O(d) \subset U(d)$. Clearly $\Sigma_1$ contains both
Werner and isotropic states and, therefore, it contains a convex
hull  of $\widetilde{Q}^\alpha$ and $\widetilde{P}^\alpha$:
\begin{equation}\label{}
    \Sigma_1 \supset \mbox{conv}\,
    \{\widetilde{Q}^0,\widetilde{Q}^1,\widetilde{P}^0,\widetilde{P}^1\}\
    .
\end{equation}
 It is easy to see that these four states are co-planar,
i.e. they belong to a common 2-dimensional plane in
$d^2$-dimensional space of hermitian operators in $\Cd\ot \Cd$.
Indeed, one shows that
\begin{equation}\label{}
    \mbox{det} \left[ \begin{array}{c|c}
    \mbox{Tr}(\widetilde{Q}^\alpha\widetilde{Q}^\beta) &
     \mbox{Tr}(\widetilde{Q}^\alpha\widetilde{P}^\beta) \\ \hline
      \mbox{Tr}(\widetilde{P}^\alpha\widetilde{Q}^\beta) &
       \mbox{Tr}(\widetilde{P}^\alpha\widetilde{P}^\beta)
       \end{array} \right] =0 \ ,
\end{equation}
and hence $\Sigma_1$ is 2-dimensional.  Therefore the two lines
$L$ and $L_\tau$ intersect and the point $L \cap L_\tau$ is
described by
\begin{equation}\label{}
    q = \frac 12 - \frac{1}{d(d+1)}\ ,
\end{equation}
and
\begin{equation}\label{}
 p =
    \frac{2}{d(d+1)}\left[ \frac 12 + \frac{1}{d(d+1)}\right]\ .
\end{equation}
Note that $q,p \in [0,1]$ and hence the intersection point $L \cap
L_\tau \in \Sigma_1$ defines a state which is both Werner and
isotropic. Moreover, since $q< 1/2$  (and $p< 1/d$) this state is
separable.

Now, it turns out that $\Sigma_1$ defines a simplex with vertices
$\widetilde{\Pi}^\alpha; \ \alpha=0,1,2$, where
\begin{eqnarray}\label{Pi}
\Pi^0 &=& Q^0 - P^1 \  , \nonumber \\
\Pi^1 &=& Q^1\ , \\
\Pi^2 &=& P^1\ . \nonumber
\end{eqnarray}
One may call it a 'minimal simplex' containing $\mbox{conv}\,
    \{\widetilde{Q}^0,\widetilde{Q}^1,\widetilde{P}^0,\widetilde{P}^1\}$. In particular
\begin{equation}\label{}
    \widetilde{Q}^0 = \frac{1}{d(d+1)}\, \left[  (d-1)(d+2)\,
    \widetilde{\Pi}^0 + 2\widetilde{\Pi}^2 \right]\ ,
\end{equation}
and
\begin{equation}\label{}
    \widetilde{P}^0 = \frac{1}{2(d+1)}\, \left[  (d+2)\,
    \widetilde{\Pi}^0 + d\widetilde{\Pi}^1 \right]\ .
\end{equation}

 Note, that the family $\Pi^k$ gives rise to the orthogonal resolution of
identity in $\mathcal{H}_A \ot \mathcal{H}_B$:
\begin{equation}\label{}
    \Pi^i \Pi^j = \delta_{ij}\Pi^j \ ,
\end{equation}
and
\begin{equation}\label{}
    \Pi^0 + \Pi^1 + \Pi^2 = I^{\ot 2} \ .
\end{equation}
Any  state $\rho$ in $\Sigma_1$ may be written as follows
\begin{equation}\label{rho-pi}
    \rho = \sum_{k=0}^2\, \pi_k\, \widetilde{\Pi}^k\ ,
\end{equation}
where $\widetilde{\Pi}^k = \Pi^k/\mbox{Tr}\Pi^k$, and the
corresponding `fidelities'
\begin{equation}\label{}
    \pi_k = \mbox{Tr}(\rho \Pi^k)\ ,
\end{equation}
satisfy $\pi_k \geq 0$ together with $\sum_k\pi_k =1$. It is
evident that an arbitrary bipartite state $\rho$ may be projected
onto the $O\ot O$--invariant subspace  by the following projection
operation $\, \mathbf{P}^{(1)} : \mathcal{P} \longrightarrow
\Sigma_1$:
\begin{equation}\label{cal-D}
    \mathbf{P}^{(1)}\rho = \sum_{k=0}^2\, \mbox{Tr}(\rho \mathbf{\Pi}^k)\, \widetilde{\Pi}^k\
    .
\end{equation}

\subsection{Separability and PPT condition}

Consider a separable state $\sigma = P_\psi \ot P_\varphi$, where
$P_x = |x\>\<x|$,  and $\psi,\varphi$ are normalized vectors in
$\Cd$. One easily finds for fidelities $\mbox{Tr}(\sigma \Pi^k)$:
\begin{eqnarray}\label{SEP}
    \pi_0 &=& \frac 12 (1 + \alpha) - \frac \beta d\ , \nonumber
    \\
    \pi_1 &=& \frac 12 (1- \alpha) \ , \\
    \pi_2 &=& \frac \beta d\ , \nonumber
\end{eqnarray}
where
\begin{equation}\label{}
    \alpha = |\<\psi|\varphi\>|^2 \ , \ \ \ \
    \beta = |\<\psi|\overline{\varphi}\>|^2 \ .
\end{equation}
Now, an arbitrary separable state is a convex combination of the
extremal product states $P_\psi \ot P_\varphi$. Noting that $0\leq
\alpha,\beta\leq 1$, the separable $O\ot O$--invariant states
satisfy
\begin{equation}\label{SEP||}
  \pi_1 \leq  \frac 1 2\ , \ \ \ \
    \pi_2 \leq \frac 1d  \ ,
\end{equation}
i.e. they combine separability conditions for Werner states $\pi_1
\leq 1/2$ and isotropic states $\pi_2 \leq 1/d$.

Now, applying a partial transposition $(\oper \ot \tau)$ to
(\ref{rho-pi}) one finds
\begin{equation}\label{}
(\oper \ot \tau)\rho = \sum_{\alpha=0}^2\,
\pi'_\alpha\,\widetilde{\Pi}^k\ ,
\end{equation}
where
\begin{equation}\label{}
    \pi'_\alpha = \sum_{\beta=0}^2\, \pi_\beta\,
    \mathbf{C}^{\beta\alpha}\ ,
\end{equation}
and $\mathbf{C}$ denotes the following $3 \times 3$ matrix:
\begin{equation}\label{C}
    \mathbf{C} = \frac{1}{2d}\, \left( \begin{array}{ccr}
d-2 & d & 2 \\ d+2& d & -2 \\ (d-1)(d+2) & -d(d-1) & 2
\end{array} \right) \ .
\end{equation}
Observe that
\begin{equation}\label{}
\sum_{\beta=0}^2\,
    \mathbf{C}^{\beta\alpha} = 1\ ,
\end{equation}
but $\mathbf{C}^{\beta\alpha}$ contains negative elements and
hence it is not a stochastic matrix. The Peres-Horodecki condition
\cite{Peres,PPT} implies $\pi'_\alpha \geq 0$ and hence
\begin{eqnarray}\label{}
    \pi_0 + \pi_1 - (d-1)\pi_2 &\geq & 0 \ , \\
    \pi_0 - \pi_1 + \pi_2 &\geq & 0 \ ,
\end{eqnarray}
which is equivalent to $\pi_1 \leq 1/2$ and $\pi_2 \leq 1/d$. This
shows that bipartite $O \ot O$--invariant state is separable  iff
it is PPT.

\section{$2 \times 2$--partite states}

\subsection{Construction}

Consider now a 4-partite system living in $\mathcal{H}_1 \ot
\mathcal{H}_2 \ot \mathcal{H}_3 \ot \mathcal{H}_4$ with
$\mathcal{H}_k = \Cd$. Following \cite{I} we may introduce two
Alices $A_k$ and two Bobs $B_k$: $A_k$ lives in $ \mathcal{H}_k$
and $B_k$ lives in $ \mathcal{H}_{2+k}$ (for $k=1,2$).

Let $\mbox{\boldmath $\alpha$}$ be a trinary $2$-dimensional
vector, i.e. $\mbox{\boldmath $\sigma$}= (\alpha_1,\alpha_2)$ with
$\alpha_j \in \{0,1,2\}$. Following \cite{I} we define a family of
$4$--partite projectors
\begin{equation}\label{Pi-2-2}
    \mathbf{\Pi}^{\sbalpha} = \Pi^{\alpha_1}_{1|3} \ot
 \Pi^{\alpha_2}_{2|4} \ ,
\end{equation}
where $L_{i|j}$ denotes a bipartite operator acting on
$\mathcal{H}_i \ot \mathcal{H}_j$, and $\Pi^\alpha$ are defined in
(\ref{Pi}). One easily shows that 9 projectors (\ref{Pi-2-2})
satisfy
\begin{enumerate}
\item $\ \ \mathbf{\Pi}^\sbalpha$ are
$\mathbf{O}\ot \mathbf{O}$--invariant, i.e.
\begin{equation}\label{}
    \mathbf{O} \ot \mathbf{O}\, \mathbf{\Pi}^\sbalpha = \mathbf{\Pi}^\sbalpha \mathbf{O}\ot
    \mathbf{O}\ ,
\end{equation}
with $\mathbf{O} = (O_1,O_2)$, and
\[ \mathbf{O}\ot \mathbf{O} = O_1 \ot  O_2 \ot  O_1  \ot O_2\ . \]

\item $\ \ \mathbf{\Pi}^\sbalpha\cdot
\mathbf{\Pi}^\sbbeta = \delta_{\sbalpha \sbbeta}\,
\mathbf{\Pi}^\sbbeta$,

\item $\ \ \sum_{\sbalpha}\, \mathbf{\Pi}^\sbalpha\, = \,
\oper^{\ot 4}\ , $

\end{enumerate}
that is, $\mathbf{\Pi}^\sbalpha$ define spectral resolution of
identity in $(\mathbb{C}^d)^{\ot 4}$. Hence, any 4-partite
$\mathbf{O}\ot \mathbf{O}$--invariant state may be uniquely
represented by
\begin{equation}\label{}
    \rho = \sum_{\sbalpha} \, \pi_\sbalpha\,
    \widetilde{\mathbf{\Pi}}^\sbalpha\ ,
\end{equation}
where the corresponding `fidelities' $\pi_\sbalpha =
\mbox{Tr}(\rho\, {\mathbf{\Pi}}^\sbalpha)$ satisfy $\pi_\sbalpha
\geq 0$ together with $\sum_\sbalpha \pi_\sbalpha = 1$. The above
construction gives rise to 8-dimensional simplex $\Sigma_2$ with
vertices $\widetilde{\mathbf{\Pi}}^\sbalpha$. Note, that
$\Sigma_2$ contains a convex hull of 4 classes of 4-partite
invariant states introduced in \cite{I}:
\begin{equation}\label{}
    \Sigma_2 \supset \mbox{conv}\, \Big\{
    \Sigma^{(00)}_2,\Sigma^{(01)}_2,\Sigma^{(10)}_2,\Sigma^{(11)}_2 \Big\}
    \ ,
\end{equation}
where
\begin{eqnarray}\label{}
    \Sigma^{(00)}_2 &=& \mbox{conv} \, \{ \widetilde{Q}^{i}_{1|3} \ot
    \widetilde{Q}^{j}_{2|4} \} \ , \\
\Sigma^{(01)}_2 &=& \mbox{conv} \, \{ \widetilde{Q}^{i}_{1|3} \ot
    \widetilde{P}^{j}_{2|4} \} \ ,\\
    \Sigma^{(10)}_2 &=& \mbox{conv} \, \{ \widetilde{P}^{i}_{1|3} \ot
    \widetilde{Q}^{j}_{2|4} \} \ ,\\
    \Sigma^{(11)}_2 &=& \mbox{conv} \, \{ \widetilde{P}^{i}_{1|3} \ot
    \widetilde{P}^{j}_{2|4} \} \ ,
\end{eqnarray}
with $i,j \in \{0,1\}$. A 3-dimensional simplex $\Sigma^\sba_2$,
where $\ba=(a_1,a_2)$ denotes 2-dimensional binary vector, defines
a set of $\ba$-invariant states. Recall that a 4-partite state
$\rho$ is $\ba$--invariant iff $\tau_\sba \rho$, with
\begin{equation}\label{tau-a-2}
\tau_\sba = \oper \ot \oper \ot \tau^{a_1} \ot \tau^{a_2}\ ,
\end{equation}
 is $\mathbf{U} \ot \mathbf{U}$--invariant. In particular
$\Sigma^{(00)}_2$ and $\Sigma^{(11)}_2$ denote 4-partite Werner
and isotropic states, respectively.

\subsection{Separability}

To find the corresponding separability criteria note that a
general 4-partite $O\ot O$--invariant state $\rho$ is 4-separable
iff there exists a 4-separable state $\sigma$ such that
$\mathbf{P}^{(2)}\rho = \sigma$, where
\begin{equation}\label{}
\mathbf{P}^{(2)} : \mathcal{P} \longrightarrow \Sigma_2\ ,
\end{equation}
defines a projection onto 4-partite $O\ot O$--invariant states.
 Consider an extremal
product state $\sigma = P_{\psi_1} \ot P_{\psi_2} \ot
P_{\varphi_1} \ot P_{\varphi_2}$, where $\psi_i,\varphi_j$ are
normalized vectors in $\Cd$. One easily finds for fidelities
$\mbox{Tr}(\sigma \Pi^\sbsigma)$:
\begin{eqnarray}\label{pi-sep}
    \pi_\sbsigma &=& \mbox{Tr}(P_{\psi_1} \ot P_{\varphi_1}\cdot
    \Pi_{1|3}^{\sigma_1} )\, \mbox{Tr}(P_{\psi_2} \ot P_{\varphi_2}\cdot
    \Pi_{2|4}^{\sigma_2} )\nonumber \\ &=& u_1\cdot u_2\ ,
\end{eqnarray}
with
\begin{equation}\label{ui}
    u_i = \left\{ \begin{array}{ll}
    (1 + \alpha_i)/2 - \beta_i/d \  \ \ &\ , \ \ \sigma_i =0 \\
    (1- \alpha_i)/2 & \ ,\ \ \sigma_i = 1 \\
    \beta_i/d & \ ,\ \ \sigma_i = 2 \end{array} \right. \ ,
\end{equation}
where
\begin{equation}\label{}
    \alpha_i = |\<\psi_i|\varphi_i\>|^2 \ , \ \ \ \
    \beta_i = |\<\psi_i|\overline{\varphi}_i\>|^2 \ .
\end{equation}
Now, since $\alpha_i,\beta_i \leq 1$, the projection
$\mathbf{P}^{(2)}$ of the convex hull of extremal separable states
gives the subset of separable $O\ot O$--invariant states defined
by the following relations:
\begin{equation}\label{pi-II}
    \pi_\sbsigma \leq \frac{1}{f_{\sigma_1}f_{\sigma_2}}\ ,
\end{equation}
where
\begin{equation}\label{ff}
    f_\sigma = \left\{ \begin{array}{ll} 1  & ,\ \sigma=0 \\
    2  & ,\ \sigma=1 \\ d  &, \   \sigma=2 \end{array} \right.
    \ .
\end{equation}
It is evident, that (\ref{pi-II}) generalize formulae
(\ref{SEP||}). Clearly,  separable states in $\Sigma_2$ contain a
convex hull of separable states in each $\ba$--invariant simplex
$\Sigma^\sba_2$:
\begin{equation}\label{}
    \mbox{Sep}(\Sigma_2) \supset \mbox{conv}\, \bigcup_{\sba}\,
    \mbox{Sep}(\Sigma^\sba_2)\ .
\end{equation}
Is 4-separability equivalent to PPT condition? Note, that one may
perform 3 different partial transpositions (\ref{tau-a-2}):
\begin{eqnarray}\label{}
    \tau_{(01)} &=& \oper \ot \oper \ot \oper \ot \tau\ ,\nonumber
    \\
    \tau_{(10)} &=& \oper \ot \oper \ot \tau \ot \oper\ , \\
    \tau_{(11)} &=& \oper \ot \oper \ot \tau \ot \tau\ . \nonumber
\end{eqnarray}
It is easy to see that
\begin{eqnarray}\label{01-PPT}
\tau_{(01)}\, \rho &=& \sum_\sbalpha\, \pi'_\sbalpha\,
\widetilde{\mathbf{\Pi}}^\sbalpha\ , \\
\label{10-PPT}\tau_{(10)}\, \rho &=& \sum_\sbalpha\,
\pi''_\sbalpha\, \widetilde{\mathbf{\Pi}}^\sbalpha\ , \\
\label{11-PPT}\tau_{(11)}\, \rho &=& \sum_\sbalpha\,
\pi'''_\sbalpha\, \widetilde{\mathbf{\Pi}}^\sbalpha\ ,
\end{eqnarray}
with
\begin{eqnarray}\label{pi-01}
 \pi'_\sbalpha   & =& \sum_\sbbeta\, \pi_\sbbeta\,(\mathbf{I}\ot
 \mathbf{C})^{\sbbeta\sbalpha}\ , \\ \label{pi-10}
 \pi''_\sbalpha   & =& \sum_\sbbeta\, \pi_\sbbeta\,(\mathbf{C}\ot
 \mathbf{I})^{\sbbeta\sbalpha}\ , \\ \label{pi-11}
 \pi'''_\sbalpha   & =& \sum_\sbbeta\, \pi_\sbbeta\,(\mathbf{C}\ot
 \mathbf{C})^{\sbbeta\sbalpha}\ ,
\end{eqnarray}
where $\mathbf{I}$ denotes $3\times 3$ identity matrix and
$\mathbf{C}$ is defined in (\ref{C}). For example one finds that a
state $\rho \in \Sigma_2$ is $(01)$--PPT, i.e. $\tau_{01}\rho \geq
0$ iff
\begin{eqnarray}\label{A01-PPT}
 \pi_{00} + \pi_{01} - (d-1)\pi_{02} &\geq 0\ ,\nonumber \\
\pi_{00} -\pi_{01} + \pi_{02} &\geq 0\ ,\nonumber\\
 \pi_{10} + \pi_{11} - (d-1)\pi_{12} &\geq 0\ ,\\
\pi_{10} - \pi_{11} + \pi_{12} &\geq 0\ ,\nonumber\\
\pi_{20} + \pi_{21} - (d-1)\pi_{22} &\geq 0\ ,\nonumber\\
 \pi_{20} -\pi_{21} + \pi_{22} &\geq 0\ . \nonumber
\end{eqnarray}
Similarly, it is $(10)$--PPT iff
\begin{eqnarray}\label{A10-PPT}
 \pi_{00} + \pi_{10} - (d-1)\pi_{20} &\geq 0\ , \nonumber\\
\pi_{00} -\pi_{10} + \pi_{20} &\geq 0\ ,\nonumber\\
 \pi_{01} + \pi_{11} - (d-1)\pi_{21} &\geq 0\ ,\\
\pi_{01} - \pi_{11} + \pi_{21} &\geq 0\ ,\nonumber\\
\pi_{02} + \pi_{12} - (d-1)\pi_{22} &\geq 0\ ,\nonumber\\
 \pi_{02} -\pi_{12} + \pi_{22} &\geq 0\ . \nonumber
\end{eqnarray}
Now, it was proved in \cite{I} that any 4-partite $\mathbf{U} \ot
\mathbf{U}$--invariant state is 4-separable iff it is $(01)$-
$(10)$- and $(11)$-PPT. Moreover, any symmetric state  is $A|B$
bi-separable iff it is $(11)$--PPT. We conjecture that the same
property holds for $\mathbf{O}\ot \mathbf{O}$--invariant states.
To prove it one has to apply the same techniques as in \cite{I}.
To investigate all PPT conditions one needs  together with
(\ref{A01-PPT}) and (\ref{A10-PPT}) a highly complicated
$(11)$--PPT condition which we shall not consider here.

\section{$2K$--partite states}

\subsection{General contruction}

Generalization to $2K$--partite system is straightforward.
Following \cite{I} we introduce $2K$ qudits with the total space
$\mathcal{H} = \mathcal{H}_1 \ot \ldots \ot \mathcal{H}_{2K} =
(\mathbb{C}^d)^{\ot 2K}$. We may still interpret the total system
as a bipartite one with $\mathcal{H}_A =\mathcal{H}_1 \ot \ldots
\ot \mathcal{H}_{K}$ and $\mathcal{H}_B =\mathcal{H}_{K+1} \ot
\ldots \ot \mathcal{H}_{2K}$. Equivalently, we may introduce $K$
Alices and $K$ Bobs with $\mathcal{H}_{A_i} = \mathcal{H}_i$ and
$\mathcal{H}_{B_i} = \mathcal{H}_{K+i}$, respectively. Then
$\mathcal{H}_A$ and $\mathcal{H}_B$ stand for the composite $K$
Alices' and Bobs' spaces.

Now, let $\mbox{\boldmath $\alpha$}$ be a trinary $K$-dimensional
vector, i.e. $\mbox{\boldmath $\sigma$}=
(\alpha_1,\ldots,\alpha_K)$ with $\alpha_j \in \{0,1,2\}$. In
analogy to (\ref{Pi-2-2}) let us define a family of $2K$--partite
projectors
\begin{equation}\label{}
    \mathbf{\Pi}^{\sbalpha} = \Pi^{\alpha_1}_{1|K+1} \ot \ldots
    \ot \Pi^{\alpha_K}_{K|2K} \ .
\end{equation}
One easily shows that
\begin{enumerate}
\item $\ \ \mathbf{\Pi}^\sbalpha$ are
$\mathbf{O}\ot \mathbf{O}$--invariant, i.e.
\begin{equation}\label{}
    \mathbf{O} \ot \mathbf{O} \, \mathbf{\Pi}^{\sbalpha} =
    \mathbf{\Pi}^{\sbalpha} \mathbf{O}\ot
    \mathbf{O}\ ,
\end{equation}
with $\mathbf{O} = (O_1,\ldots,O_K)$, and
\[ \mathbf{O}\ot \mathbf{O} = O_1 \ot \ldots \ot O_K \ot  O_1 \ot
\ldots \ot O_K\ . \]

\item $\ \ \mathbf{\Pi}^\sbalpha\cdot
\mathbf{\Pi}^\sbbeta = \delta_{\sbalpha \sbbeta}\,
\mathbf{\Pi}^\sbbeta$,

\item $\ \ \sum_{\sbalpha}\, \mathbf{\Pi}^\sbalpha\, = \,
\oper^{\ot 2K}\ . $

\end{enumerate}
Therefore, $2K$--partite  $\mathbf{O}\ot \mathbf{O}$--invariant
states define a $(3^{K}-1)$--dimensional simplex $\Sigma_K$:
\begin{equation}\label{}
    \rho = \sum_{\sbalpha} \, \pi_\sbalpha\,
    \widetilde{\mathbf{\Pi}}^\sbalpha\ ,
\end{equation}
where
\begin{equation}\label{}
    \widetilde{\mathbf{\Pi}}^{\sbalpha} = \widetilde{\Pi}^{\alpha_1}_{1|K+1} \ot \ldots
    \ot \widetilde{\Pi}^{\alpha_K}_{K|2K} \ ,
\end{equation}
 and the corresponding `fidelities'
\begin{equation}\label{}
    \pi_\sbalpha = \mbox{Tr}(\rho \Pi^\sbalpha)\ ,
\end{equation}
satisfy $\pi_\sbalpha \geq 0$ together with
$\sum_\sbalpha\pi_\sbalpha =1$.

Denote by $\Sigma_K^\sba$ a $(2^K-1)$--dimensional simplex of
$\!\ba$--invariant states, where $\ba=(a_1,\ldots,a_K)$ denotes a
binary $K$-vector. Recall that a $2K$-partite state $\rho$ is
$\ba$--invariant iff $\tau_\sba \rho$, with
\begin{equation}\label{tau-a}
\tau_\sba = \oper^{\ot K} \ot \tau^{a_1} \ot \ldots \ot
\tau^{a_K}\ ,
\end{equation}
 is $\mathbf{U} \ot \mathbf{U}$--invariant. In particular
$\Sigma^{(0\ldots 0)}_K$ and $\Sigma^{(1\ldots 1)}_K$ denote the
simplex of $2K$-partite Werner and isotropic states, respectively
(see \cite{I}). It is therefore clear that $\Sigma_K$ contains a
convex hull of $2^K$ single $\ba$-invariant simplexes
$\Sigma^\sba_K$:
\begin{equation}\label{}
    \Sigma_K \supset \mbox{conv}\, \bigcup_{\sba}\,
    \Sigma^\sba_K\ .
\end{equation}

\subsection{Separability and multi--PPT conditions}

To find separability conditions for $2K$-partite $\mathbf{O}\ot
\mathbf{O}$--invariant states consider a separable state
\[ \sigma = P_{\psi_1} \ot \ldots \ot
P_{\psi_K} \ot  P_{\varphi_1} \ot \ldots \ot P_{\varphi_K}\ , \]
where $\psi_i,\varphi_j$ are normalized vectors in $\Cd$. One
easily finds for fidelities $\mbox{Tr}(\sigma \Pi^\sbsigma)$:
\begin{eqnarray}\label{pi-sepK}
    \pi_\sbsigma &=& \prod_{i=1}^K\, \mbox{Tr}(P_{\psi_i} \ot P_{\varphi_i}\cdot
    \Pi_{i|K+i}^{\sigma_i} )\nonumber \\ &=& u_1\, \ldots\,  u_K\ ,
\end{eqnarray}
where $u_i$ are defined in (\ref{ui}). The projection
$\mathbf{P}^{(K)}$ of the convex hull of extremal separable states
gives the subset of separable $O\ot O$--invariant states defined
by the following relations:
\begin{equation}\label{pi-2K}
    \pi_\sbsigma \leq \frac{1}{f_{\sigma_1}\ldots f_{\sigma_K}}\ ,
\end{equation}
where $f$'s are defined in (\ref{ff}). Clearly, a set of separable
states in $\Sigma_K$ contains a convex hull of separable states in
each $\ba$--invariant simplex $\Sigma^\sba_K$:
\begin{equation}\label{}
    \mbox{Sep}(\Sigma_K) \supset \mbox{conv}\, \bigcup_{\sba}\,
    \mbox{Sep}(\Sigma^\sba_K)\ .
\end{equation}
For $2K$--partite state one may look for $2^K-1$ partial
transpositions
\begin{equation}\label{}
    \tau_\sba = \oper^{\ot K} \ot \tau^{a_1} \ot \ldots \ot
    \tau^{a_K} \ .
\end{equation}
Note, that
\begin{equation}\label{}
 \tau_\sba \rho = \sum_\sbalpha\, \pi'_\sbalpha\,
 \widetilde{\mathbf{\Pi}}^\sbalpha\ ,
\end{equation}
with
\begin{equation}\label{}
     \pi'_\sbalpha   = \sum_\sbbeta\, \pi_\sbbeta\,(\mathbf{C}^{a_1}
     \ot\ldots \ot
 \mathbf{C}^{a_K})^{\sbbeta\sbalpha}\ ,
\end{equation}
where
\begin{equation}\label{}
    {\bf C}^a \ = \ \left\{ \begin{array}{ll} \mathbf{I}\ \ &\ , \
    \ a=0 \\ \mathbf{C}\ \ &\ , \
    \ a=1 \end{array} \right. \ .
\end{equation}
In analogy to 4-partite symmetric states we conjecture that a
$2K$-partite state in $\Sigma_K$ is $2K$-separable iff it is
$\bb$-PPT for all binary 2-vectors $\bb$. Moreover, a state in
$\Sigma_K$ is $A|B$ bi-separable iff it is $(1\ldots 1)$--PPT.

\subsection{Reductions}

It is evident that reducing the $2K$ partite state  $\rho \in
\Sigma_K$ with respect to $A_i \ot B_i$ pair one obtains
$2(K-1)$--partite state  $\rho' \in \Sigma_{K-1}$ living in
\begin{equation}\label{}
    \mathcal{H}_1 \ot \ldots \check{\mathcal{H}}_i \ot \ldots \ot
    \check{\mathcal{H}}_{i+K} \ot \ldots \ot \mathcal{H}_{2K}\ ,
\end{equation}
where $\check{\mathcal{H}}_i$ denotes the omitting of
${\mathcal{H}}_i$. The corresponding fidelities are given by
\begin{equation}\label{}
   \pi'_{(\alpha_1\ldots\alpha_{K-1})}\ = \
    \sum_{\beta}\,
    \pi_{(\alpha_1\ldots\alpha_{i-1}\beta\alpha_i\ldots\alpha_{K-1})}\
    .
\end{equation}
Note, that reduction with respect to a `mixed' pair, say $A_i \ot
B_j$ with $i\neq j$, is equivalent to two `natural' reductions
with respect to $A_i \ot B_i$ and $A_j \ot B_j$ and hence it gives
rise to $2(K-2)$--partite invariant state. This procedure
establishes a natural hierarchy of multipartite  $\mathbf{O}\ot
\mathbf{O}$--invariant states.

 \acknowledgments  This work was partially supported by the
Polish State Committee for Scientific Research Grant {\em
Informatyka i in\.zynieria kwantowa} No PBZ-Min-008/P03/03.

\end{document}